\documentclass{revtex4}

\usepackage{psfig}

\begin{document}

\title{Stability of hexagonal solidification patterns}
\author{Mathis Plapp and Marcus Dejmek}
\affiliation{
   Laboratoire de Physique de la Mati\`ere Condens\'ee, \\
             CNRS/Ecole Polytechnique, 91128 Palaiseau, France
}

\date{\today}

\begin{abstract}
We investigate the dynamics of cellular solidification
patterns using three-dimensional phase-field 
simulations. The cells can organize into stable 
hexagonal patterns or exhibit unsteady evolutions. 
We identify the relevant secondary instabilities
of regular hexagonal arrays and find that the stability 
boundaries depend significantly on the strength of
crystalline anisotropy. We also find multiplet states 
that can be reached by applying well-defined 
perturbations to a pre-existing hexagonal array.
\end{abstract}

\maketitle

\section{Introduction}
Hexagonal patterns arise in many non-equilibrium 
systems when a two-dimensional
translational symmetry is broken by a dynamic 
instability \cite{CrossHoh}. Well-known examples are 
B{\'e}nard-Marangoni convection or Faraday surface waves.
In directional solidification, where alloy samples are pulled 
with a fixed speed $V$ in an externally imposed temperature 
gradient $G$, the planar solidification front becomes
unstable above a critical value of the ratio $V/G$.
Close above this threshold, the front develops shallow
cells that often form regular hexagonal arrays \cite{Rutter53,Tiller56}; 
as $V/G$ increases, cells deepen and eventually
form dendrites with lateral branches. Besides the 
interest of this sequence of morphological transitions 
as a model system for the physics of pattern formation \cite{BegRohu}, 
it also has practical importance because the
morphology of the solidification front determines 
the spatial distribution of alloy components and 
defects in the solidified material that, in turn,
influence its mechanical properties.

We focus here on two issues: the
influence of crystalline anisotropy on
pattern selection, and the existence of
multiplet states consisting of several 
symmetry-broken cells. These questions have
been examined in two dimensions (2D), both 
in thin-sample experiments \cite{Jamgot93,Akamatsu98,Losert98} 
and numerically by boundary integral and phase-field 
methods \cite{Losert98,Kopcz96,Kopcz97}.
The situation is less clear in three dimensions (3D),
where the analysis of experimental data must
account for the presence of convection in
the bulk liquid, and numerical investigations
have so far been restricted to rather short 
length and time scales \cite{Grossmann93,Abel97}.
Here, we use an efficient formulation 
of the phase-field method \cite{Karma98}
and a multi-scale simulation algorithm \cite{Plapp00}
to perform extensive quantitative 3D
simulations of a symmetric model of directional
solidification.

The influence of crystallographic effects on
solidification patterns was already noticed
in early experimental studies \cite{Rutter53,Tiller56}.
It is by now well established that for freely
growing dendrites, pattern selection is
governed by small anisotropies of the
solid-liquid surface tension and interface
kinetics. More surprisingly, it was recently shown
that anisotropy plays a crucial role even close
above the instability threshold in directional
solidification, where the cell shapes
are round \cite{Akamatsu98,Kopcz96}.
For thin-sample experiments, where
the system is quasi-two-dimensional, the
anisotropy of the one-dimensional interface is
controlled by the orientation of the crystal with
respect to the sample plane \cite{Akamatsu98}. 
Stable and regular steady states are obtained only 
for sufficiently strong anisotropy; for weak anisotropy, 
cells may split in two or disappear, and the solidification
front never reaches steady state. In numerical simulations
without anisotropy, cells are stable only in a narrow 
parameter range close to the onset of the primary instability;
when anisotropy is included, stable solutions 
exist even far from threshold \cite{Kopcz96}. Here, we 
investigate in detail the stability of three-dimensional
hexagonal cells and identify the relevant secondary
instabilities. We find that the anisotropy 
has the same strong effect on array stability 
as in 2D.

Furthermore, we find triplet solutions; that is, 
groups of three cells that grow with their 
tips close together. This is the 3D equivalent 
of doublet fingers found in 2D both in
experiments and simulations \cite{Jamgot93,Kopcz97,Losert98}.
Similar 3D growth structures have been found in 
simulations of free growth in a channel \cite{Abel97}, 
and as transients in experiments \cite{Noel97}.
In our simulations, triplets can be generated 
in a controlled manner starting from a
pre-existing hexagonal array.

\section{Model and implementation}
An alloy of overall composition $c_0$ is pulled
with velocity $V$ in a linear temperature field
with its gradient $G$ directed along the $z$ axis.
In the frame of the sample, the temperature is
given by $T=T_0 + G(z-Vt)$,
where solidification starts at $t=0$ from a flat
equilibrium interface of temperature $T_0$ at $z=0$.
This is the so-called frozen temperature approximation,
valid for slow solidification and similar thermal
conductivities of both phases.
The phase field $\phi$ distinguishes between
solid ($\phi=1$) and liquid ($\phi=-1$). The
concentration jump between solid and liquid $\Delta c$ 
is assumed constant. Instead of the concentration
field, which exhibits rapid variations across
the interface, the dimensionless field
$u = (c -c_0)/\Delta c - (1+\phi)/2$ is used.
It is constant for an equilibrium interface and 
hence equivalent to a chemical potential.
The time evolution of the fields is given by
\begin{equation}
\tau(\hat{n}) \partial_t \phi = - {\frac{\delta F[\phi,u]}{\delta \phi}},
\end{equation}
\begin{equation}
\partial_t u = D\nabla^2 u + \partial_t \phi/2,
\end{equation}
where $\hat{n} = \nabla \phi/|\nabla\phi|$ is the unit
normal to the interface, $\tau(\hat{n})$ is an 
orientation-dependent relaxation time, $D$ is the
solute diffusivity, and $\delta F/\delta \phi$ denotes
the functional derivative of the Lyapounov functional
\begin{equation}
F= \int_{V} {\frac{W(\hat{n})^2}{2}}(\nabla \phi)^2 + f(\phi)
    + \alpha g(\phi) \left(u-\frac{z-Vt}{l_T}\right),
\end{equation}
where $W(\hat{n})$ is the orientation-dependent
interface thickness, $l_T = m\Delta c/G$ is the thermal 
length with $m$ being the liquidus slope in the phase diagram,
and $f(\phi)=-\phi^2/2+\phi^4/4$ is a double-well potential.
The quantity $u-(z-Vt)/l_T$ is the local supersaturation,
and its product with $g(\phi)=\phi-2\phi^3/3+\phi^5/5$ 
introduces a free energy difference between the two 
phases by tilting the double well, 
which provides the driving force for the phase transformation.
Anisotropic surface tension with cubic 
symmetry and one principal axis aligned with the growth 
direction is implemented by $W(\hat{n}) = W_0 a(\hat{n})$ with 
$a(\hat{n})=(1-3\epsilon_4)[(1+4\epsilon_4(n_x^4+n_y^4+n_z^4)/(1-3\epsilon_4)]$.
In the thin-interface limit \cite{Karma98}, this
model is equivalent to the classic free-boundary problem of
solidification, in which $u$ satisfies the diffusion equation 
in solid and liquid, the Stefan condition at the interface, 
$V_n = \hat{n}\cdot[D\nabla u|_s - D\nabla u|_l]$, where
$V_n$ is the normal velocity of the interface, and the
gradients of $u$ are evaluated on the solid and liquid
side of the interface, and the generalized Gibbs-Thomson
condition at the interface,
\begin{equation}
u_{\rm int} = - \frac{z_{\rm int}-Vt}{l_T} - d_0 \sum_{i=1}^2\left[a(\hat{n})
+\partial_{\theta_i}^2a(\hat{n})\right]\kappa_i - \beta(\hat{n}) V_n,
\end{equation}
where $\theta_i$ are the local angles between $\hat{n}$ and 
the two local principal directions on the interface, $\kappa_i$ 
are the principal curvatures, $d_0 = a_1 W_0/\alpha$ is the
capillary length, and 
$\beta(\hat{n})=a_1 \tau/(W\alpha)[1-a_2\alpha W^2/(D\tau)]$
is the kinetic coefficient, with $a_1=5\sqrt{2}/8$ and $a_2=0.6267$.
Taking into account grid 
corrections as explained in detail in Ref.~\cite{Karma98},
the interface kinetics can be eliminated ($\beta=0$) by setting
$\tau(\hat{n}) = \tau_0 (1-3\delta)[(1+4\delta(n_x^4+n_y^4+n_z^4)/(1-3\delta)]$
and by choosing specific values for the constants
$\tau_0$, $\delta$, and $\alpha$ that depend on $D$, the
capillary anisotropy $\epsilon_4$ and the discretization;
the values used for our simulations are listed in table~\ref{params}.

%-------------------table-----------------------
%  table of simulation parameters
\begin{table}
\caption{Simulation parameters. 
For all simulations, $W_0=1$, $D=2$, the
grid spacing is $\Delta x$=0.8 and the 
time step $\Delta t=0.05$. To obtain a ``physical'' 
capillary anisotropy $\epsilon_4$, a slightly
higher value $\epsilon_4'$ has to be used in
the model to correct for grid effects 
(see Ref.~\protect\cite{Karma98}).}
\label{params}
\begin{center}
\begin{tabular}{ccccc}
$\epsilon_4$ & $\epsilon_4'$ & $\alpha$ & $\tau_0$ & $\delta$ \\
0.00 & 0.0029 & 3.482 & 1.018 & -0.0180 \\
0.01 & 0.0129 & 3.407 & 0.9965 & 0.00345 \\
0.02 & 0.0231 & 3.330 & 0.975 & 0.0255 \\
0.03 & 0.0334 & 3.262 & 0.955 & 0.0472 \\
%0.05 & 0.0529 & 3.131 & 0.918 & 0.0890 \\
\end{tabular}
\end{center}
\end{table}
%------------------------------------------------

The involved physical length scales are the thermal
length $l_T$, the diffusion length $l_D=D/V$, and the
capillary length $d_0$. The stability of a planar
interface is controlled by the parameter $\nu=l_T/l_D$.
In experiments, $\nu$ is usually varied by changing
$V$ at fixed $G$; in the simulations, it is more efficient 
to vary $G$, and hence $l_T$ at fixed $d_0/l_D$.
Typical experimental values for $d_0/l_D$ are of
the order $10^{-4}$. However, due to computational
constraints, such small values are too costly to
simulate. Indeed, convergence of the phase-field
method requires $W/l_D\ll 1$ and $V\tau/W < 1$.
The latter, combined with the requirement
$\beta=0$ (that is, $D\tau/W^2\sim\alpha\sim W/d_0$),
yields $W/l_D < \sqrt{d_0/l_D}$. Since the discretization
is on the scale of $W$, whereas typical cell sizes
are of order $l_D$ or larger, the required number 
of grid points increases when $d_0/l_D$ decreases.
For the value $d_0/l_D=2.5\times 10^{-2}$ used in 
most of our simulations, the morphological
instability occurs at $\nu_c=3.75$ (instead of
$\nu_c=2$ in the limit $d_0/l_D\to 0$).

The above model has parallel liquidus and solidus 
lines and equal diffusivities in the solid and the 
liquid, whereas alloys exhibit strongly
different diffusivities in the two phases and a
temperature-dependent concentration jump. Nevertheless, 
this model has been shown to reproduce even complex
qualitative features of pattern selection that are
seen in 2D experiments \cite{Losert98}, and computationally
remains the most efficient model that captures
the salient features of directional solidification,
despite recent advances on more realistic alloy
models \cite{Karma01}.

\section{Influence of anisotropy}
\begin{figure}
\centerline{
\psfig{file=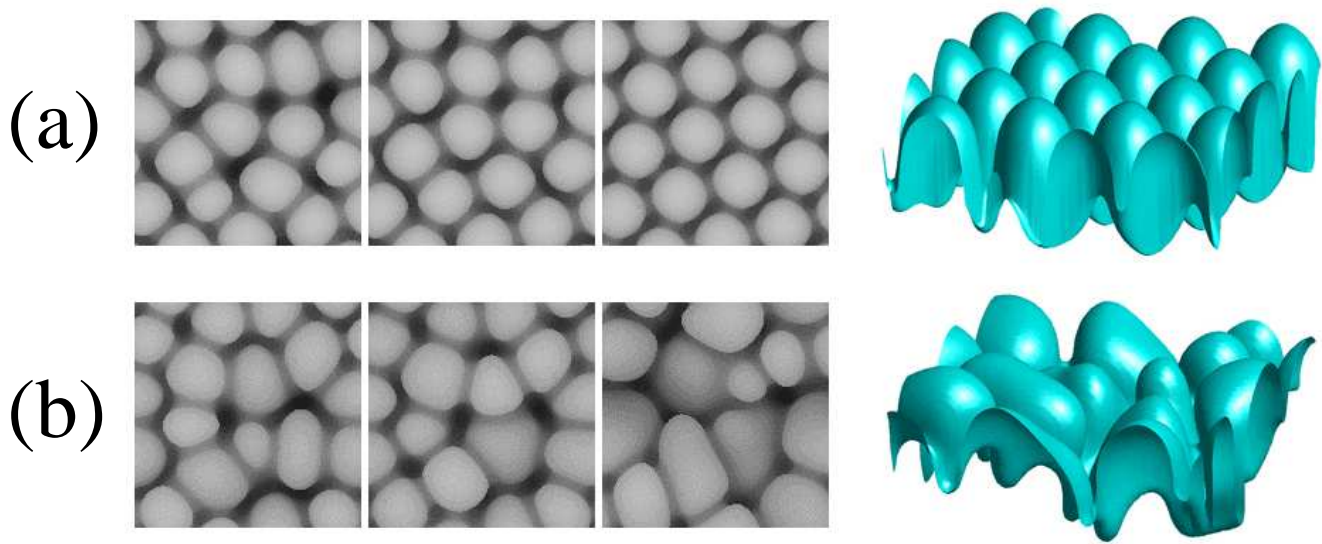,width=\textwidth}}
\caption{Evolution of cellular patterns for $\nu=12.5$ and
(a) $\epsilon_4=0.03$ and (b) $\epsilon_4=0.02$. 
Left: top views of the growth front at $tD/l_D^2=320$, $640$, and 
$1996.8$ (the greyscale is proportional to the surface height;
box size: $L_x/l_D=L_y/l_D=12.8$);
right: 3D view of the interface at the end of the run.}
\label{fig1}
\end{figure}

Simulations begin with an unstable
planar steady-state interface, and numerical noise
triggers the morphological instability. 
The subsequent evolution of the front dramatically 
depends on the anisotropy strength. For an 
anisotropy of $\epsilon_4=0.03$, some of the initial 
cells are eliminated, but most survive, become 
stable, and slowly drift over small distances 
until they form a perfectly regular lattice (Fig.~\ref{fig1}a).
In other runs, hexagonal arrays with defects are 
formed; in all cases, the interface shape 
converges to a stable steady-state solution. 
In contrast, for a slightly lower 
anisotropy ($\epsilon_4=0.02$) with otherwise
identical conditions, the cells do not reach
steady-state, and cell-elimination and cell-splitting
events occur up to the end of our simulation 
runs (Fig.~\ref{fig1}b).
%--------------------figure--------------------
%----------------------------------------------

\begin{figure}
\centerline{
\psfig{file=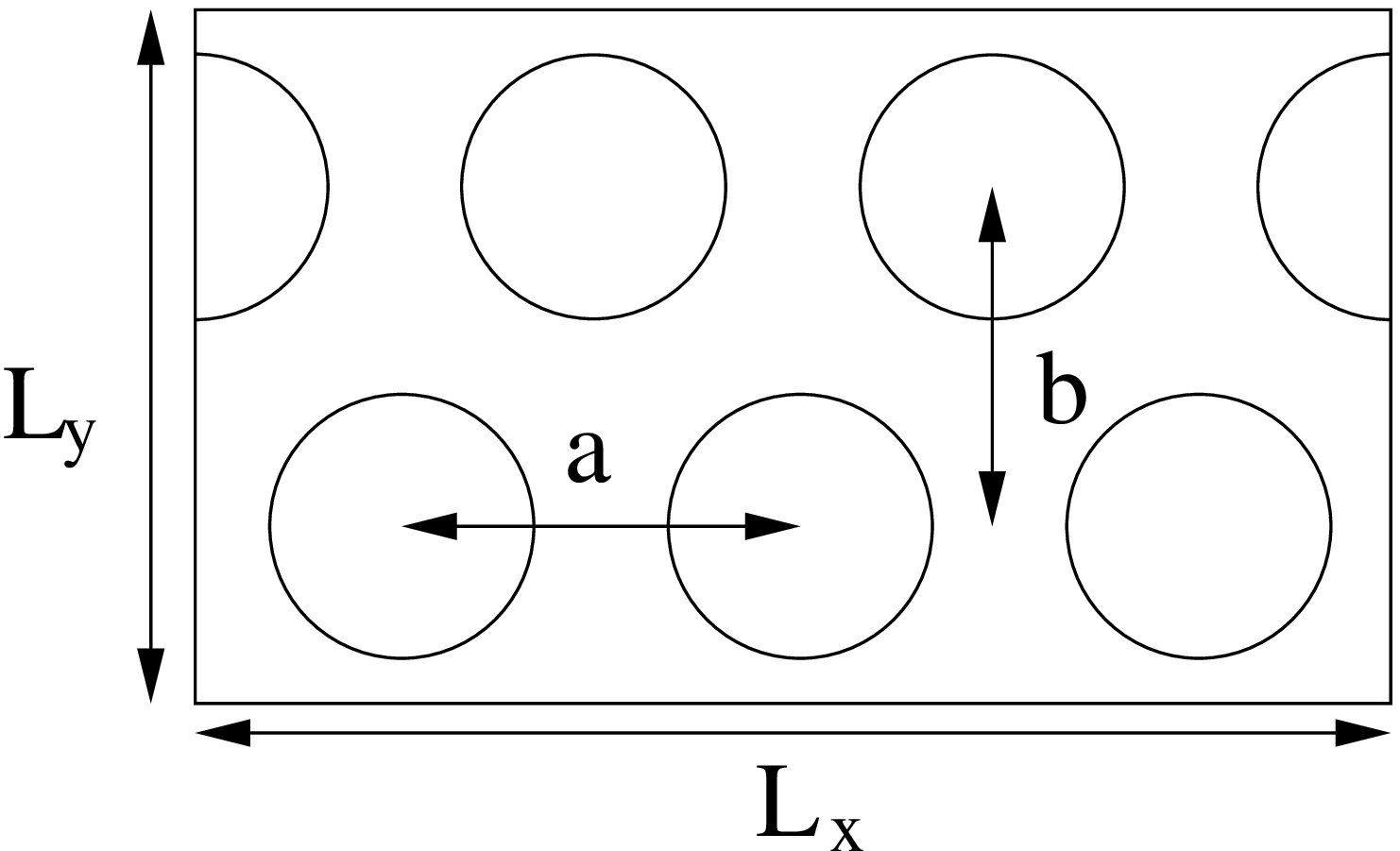,width=.4\textwidth}
\hspace{1cm}
\psfig{file=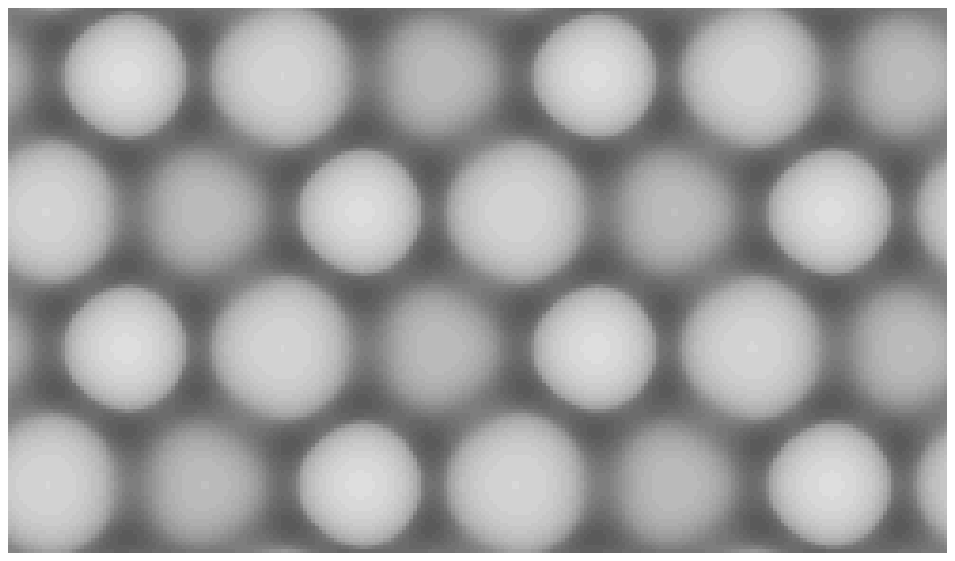,width=.4\textwidth}}
\caption{Left: sketch of the simulated geometry:
in a box of lateral size $L_x\times L_y$ with
periodic boundary conditions, $m$ rows of $n$ cells
may be generated by starting from a flat surface
perturbed by the superposition of three plane waves;
we define $a=L_x/n$ and $b=L_y/m$. Right:
snapshot of an oscillatory instability that 
develops spontaneously in a box with 24 cells ($m=4$, $n=6$) 
for $\nu=5.4$, $a/l_D=3.307$, $b/l_D=2.880$, and $\epsilon_4=0.01$.}
\label{fig2}
\end{figure}

As a first step toward the understanding of pattern 
selection, we investigate the stability of
regular hexagonal arrays.
We start from initial conditions as
described in Fig.~\ref{fig2} and allow the cells 
to reach steady state at moderate
values of $\nu$. Then, we increase $\nu$ in small
steps until a secondary instability threshold is reached. 
Two different instabilities are encountered. For large 
spacings, the cells oscillate and form a superlattice
pattern shown in Fig.~\ref{fig2}.
Three equivalent sublattices of hexagonal
symmetry emerge, with $\sqrt{3}$ times larger 
spacing than the basic pattern. All 
cells on each sublattice oscillate in phase with an
exponential growth in amplitude over time.
Eventually, the cells tip-split, and the subsequent
dynamics is unsteady and similar to Fig.~\ref{fig1}b. 
For smaller spacings, cell elimination occurs:
all cells located on the same
sublattice fall behind the others and are overgrown.
The subsequent evolution is again unsteady. Finally,
for $\nu$ very close to the onset of the primary
instability, a few examples of isolated
cell eliminations occurred that are probably the
signature of an Eckhaus instability.
%--------------------figure--------------------
%----------------------------------------------

\begin{figure}
\centerline{
\psfig{file=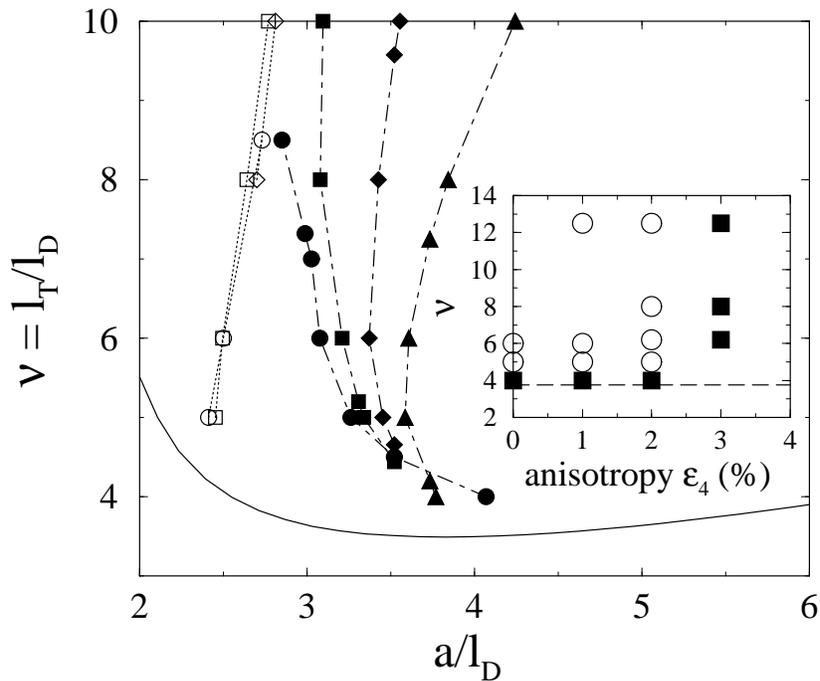,width=.6\textwidth}}
\caption{Main plot: stability boundaries for hexagonal 
arrays for different anisotropies: 0\% (circles), 
1\% (squares) 2\% (diamonds), and 3\% (triangles). 
Hexagonal arrays are stable between the
limits of cell elimination (dotted lines and open 
symbols) and oscillations (dash-dotted lines and 
full symbols). The full line is the neutral stability 
line of the primary instability. Inset: behavior
of large runs started from a flat surface. Filled
squares: stable hexagons, open circles: unsteady 
evolution, dashed line: primary instability
threshold.}
\label{bounds}
\end{figure}

The geometry of the instability modes
is entirely determined by the underlying hexagonal
symmetry of the initial pattern \cite{Pirat03}. 
Similar superlattice oscillations have been 
reported in simulations of two-dimensional amplitude 
equations \cite{Daumont97,Kassner98}, and experimentally 
observed in fluid systems \cite{Wagner00,Rogers00}.
The new feature here is the presence of 
anisotropy. Figure~\ref{bounds} shows the stability 
boundaries of oscillatory and cell elimination instabilities
computed for several values of $\epsilon_4$. 
Whereas the cell elimination mode 
is only weakly sensitive to anisotropy, the critical 
cell spacing for the onset of the oscillatory mode
is strongly anisotropy-dependent. Without
anisotropy, the stability boundaries of the
oscillatory and cell elimination modes cross and
no stable states were found for $\nu>9$.
For $\epsilon_4=0.01$, a narrow range of 
stable states was found to extend to the largest
values of $\nu$ investigated; this stable band 
becomes larger with increasing anisotropy.
These findings correlate well with the
overall behavior of the large-scale runs of 
Fig.~\ref{fig1}, depicted in the inset of 
Fig.~\ref{bounds}: Stable cellular 
arrays appear spontaneously only when a sufficiently 
large band of stable states is available.
%--------------------figure--------------------
%----------------------------------------------

Remarkably, the structure of the stability diagram is 
identical to the one obtained for 2D in Ref.~\cite{Kopcz96}.
The cell elimination and cell oscillation modes found
here for hexagonal arrays are the equivalent of the 
well-known steady and oscillatory period-doubling instabilities
found in 2D systems. To further investigate this similarity, 
we have performed two-dimensional simulations 
of our phase-field model and found that, for fixed
simulation parameters, the stability diagrams in
2D and 3D are not identical; however, the
critical spacings for the onset of instability 
differ by less than 20\% in all cases investigated.

\section{Other instability modes}
\begin{figure}
\centerline{
\psfig{file=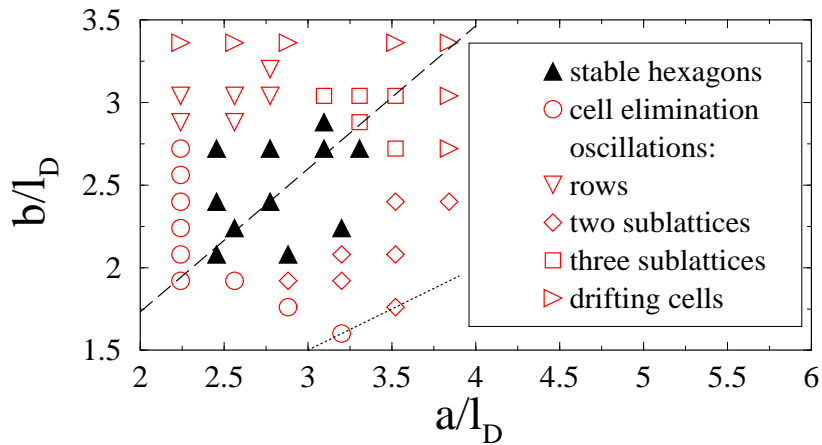,width=.6\textwidth}}
\caption{Instabilities of deformed hexagons
for $\nu=5$ and $\epsilon_4=0$. The lengths
$a$ and $b$ are defined in Fig.~\protect\ref{fig2};
the dashed line corresponds to regular hexagons, 
the dotted line to squares.}
\label{elong}
\end{figure}

To investigate how the spatial structure of
the oscillatory modes depends on the symmetry
of the underlying steady-state solution, simulations 
starting from elongated hexagons were performed.
By varying the aspect ratio of the simulation
box, patterns ranging from rows of elongated 
hexagons ($b/a > \sqrt{3}/2$) over symmetric 
hexagons ($b/a = \sqrt{3}/2$) to squares 
($b/a = 1/2$) can be generated. For a square lattice, 
there are only two equivalent sublattices,
and a two-sublattice oscillation mode emerges. 
For rows of elongated hexagons, close to a 
one-dimensional symmetry, in-phase oscillations 
of entire rows occur.
For strongly elongated hexagons, a tilt 
instability is found that corresponds to a breaking 
of the parity symmetry: the cells become asymmetric 
and drift along their longer axis. However, these 
states are unstable with respect to subsequent
oscillations. Figure~\ref{elong} shows
the symmetry of the fastest instability
in the plane of the two lengths $a$ and $b$ defined 
in Fig.~\ref{fig2}. We find a 
balloon of stable arrays consisting of regular
or slightly elongated hexagons.
%--------------------figure--------------------
%----------------------------------------------
%--------------------figure--------------------
\begin{figure}
\centerline{
\psfig{file=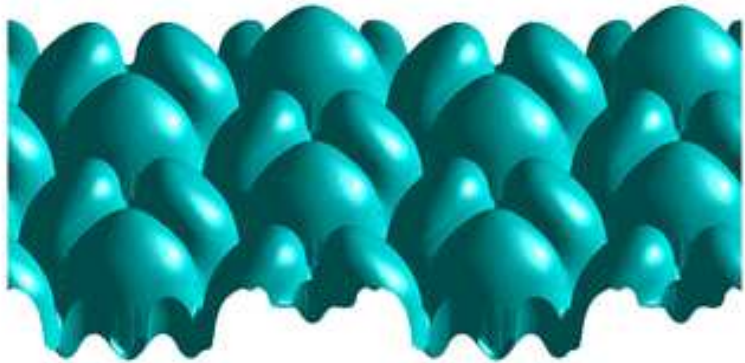,width=.6\textwidth}}
\caption{Triplets at $\nu=12.5$, $a/l_D=5.333$,
$b/l_D=4.640$, and $\epsilon_4=0.02$.
Six cells were simulated and the final picture was
replicated four times to yield a clearer view.}
\label{triplet}
\end{figure}
%----------------------------------------------

\section{Triplets}
To search for multiplet states, we use a procedure 
inspired by the experimental method used in
Ref.~\cite{Losert98} to create doublets in 
thin-sample directional solidification.
A UV-absorbing laser dye was used as solute, and by
illuminating the sample with UV light through
a mask, the temperature field was perturbed 
for a limited time with a well-defined spatial pattern.
We implement a similar procedure by adding to the
temperature field a perturbation $\Delta T(x,y)$
that has the symmetry of
one of the three sublattices of Fig.~\ref{fig2},
and whose minima are located over the ``holes'' 
between three cells. This promotes the growth of 
the surrounding cells, and their tips approach 
each other to form triplet fingers. Once the perturbation
is switched off, the system relaxes to triplet states 
that remain stable for the duration of our simulations
for large enough $\nu$ and large wavelengths 
(Fig.~\ref{triplet}). For lower $\nu$,
the triplets exhibit instabilities similar to the
usual cells, but no detailed investigation was
carried out. Triplets never appeared spontaneously
in our simulations: only unsteady evolutions resulted
from random initial conditions, even for
control parameters for which stable triplet
states exist.

\section{Conclusion}
We have studied the stability of hexagonal 
arrays and triplet fingers, and found that
the crystalline anisotropy plays a crucial role
in determining the stability bounds. Remarkably,
despite the cubic symmetry of the anisotropy,
the selected patterns are hexagonal,
while square arrays are always unstable.
Square cellular states have been calculated long ago
using a numerical scheme valid only
for shallow cells \cite{McFadden87}. However,
the geometry of the used simulation cell suppresses
all large-scale oscillatory modes, such that no
conclusion about their stability can be drawn.
Nevertheless, the existence of other than hexagonal
patterns cannot be ruled out in general, since it 
is known that the weakly nonlinear behavior of
solidification fronts strongly depends on the alloy 
characteristics \cite{BegRohu}, and we have not
carried out an exhaustive survey of the (large)
parameter space.

The main approximations made in the present
work are the constant concentration jump, the equal
solute diffusivities, and the large
value of $d_0/l_D$. The two first can
be easily levied by the use of a more general 
model \cite{Karma01}; the third remains a serious
computational challenge. We expect most of our results
to remain qualitatively valid for more realistic systems, 
since they depend only on the the underlying symmetries 
and the stability of the cell tips. In that sense, it 
is interesting to note that unsteady evolutions similar
to Fig.~\ref{fig1}b and transient isolated multiplets 
have recently been observed in 3D experiments on
transparent alloys \cite{Noel97}.

\acknowledgments
We thank Vincent Fleury for many useful discussions.
This work was supported by Centre National d'Etudes
Spatiales, France.

\end{document}